\documentclass[useASM]{mn2e}

\usepackage{graphicx}

\begin{document}

\title[Electromagnetic Emission from newly-born Magnetar Spin-Down]
{Electromagnetic Emission from newly-born Magnetar Spin-Down by Gravitational-Wave and Magnetic Dipole
Radiations}

\author[L\"{u} et al.]
{Hou-Jun L\"{u}\thanks{E-mail: lhj@gxu.edu.cn}, Le Zou, Lin Lan, En-Wei Liang\thanks{E-mail:
lew@gxu.edu.cn}\\
Guangxi Key Laboratory for Relativistic Astrophysics, School of Physical Science and Technology,
Guangxi University, Nanning 530004, China\\}
 \maketitle

\label{firstpage}
\begin{abstract}
A newly-born magnetar is thought to be central engine of some long gamma-ray bursts (GRBs). We
investigate the evolution of the electromagnetic (EM) emission from the magnetic dipole (MD) radiation
wind injected by spin-down of a newly-born magnetar via both quadrupole gravitational-wave (GW) and MD
radiations. We show that the EM luminosity evolves as $L_{\rm em}\propto (1+t/\tau_c)^{\alpha}$, and
$\alpha$ is $-1$ and $-2$ in the GW and MD radiation dominated scenarios, respectively. Transition from
the GW to MD radiation dominated epoch may show up as a smooth break with slope changing from $-1$ to
$-2$. If the magnetar collapses to a black hole before $\tau_c$, the MD radiation should be shut down,
then the EM light curve should be a plateau followed by a sharp drop. The expected generic light curve
in this paradigm is consistent with the canonical X-ray light curve of {\em Swift} long GRBs. The X-ray
emission of several long GRBs are identified and interpreted as magnetar spin-down via GW or MD, as
well as constrain the physical parameters of magnetar. The combination of MD emission and GRB
afterglows may make the diversity of the observed X-ray light curves. This may interpret the observed
chromatic behaviors of the X-ray and optical afterglow light curves and the extremely low detection
rate of a jet-like break in the X-ray afterglow light curves of long GRBs.

\end{abstract}
\begin{keywords}
star: gamma-ray burst - star: magnetar
\end{keywords}

\section {Introduction}
It is believed that gamma-ray bursts (GRBs) are resulted from collapse of massive stars or mergers of
binary compact objects, such as neutron star (NS) binaries and NS-black hole system (Zhang et al. 2007;
Kumar \& Zhang 2015 for review). Alternatively, Zhang (2006) suggested that naming the bursts as as
Type II and Type I GRBs that are consistent with the massive star origin and the compact-star-merger
origin models, respectively. The remnants of catastrophic destruction of these progenitor systems, may
be a newly-born black hole with hyper-accretion or a magnetar, serving as the central engines of GRBs
(e.g., Usov 1992; Thompson 1994; Dai \& Lu 1998a,b; Popham et al. 1999; Wheeler et al. 2000; Narayan et
al. 2001; Zhang \& M\'esz\'aros 2001; Lei et al. 2009; Metzger et al. 2011; Bucciantini et al. 2012;
L\"{u} \& Zhang 2014; Liu et al. 2017).

Theoretically, a newly-born magnetar may have different evolutional tracks. It may survive less than 1
second and quickly collapse to a black hole (Rosswog et al. 2003), or can be survived much longer, up
to a timescale of several hundreds of seconds, and even to be a long-lived NS, determining on maximum
gravitational mass and its equation of state (EOS) (Zhang \& M\'esz\'aros 2001; Zhang 2013; Giacomazzo
\& Perna 2013; Fan et al. 2013; Lasky et al. 2014; Ravi \& Lasky 2014; Metzger \& Piro 2014; Lasky \&
Glampedakis 2016; Siegel \& Ciolfi 2016). Since the injected kinetic luminosity of the magnetic dipole
(MD) radiations from a magnetar before the characteristic spin-down timescale ($\tau_{\rm c}$) is
steady, a shallow-decay segment may be observed in the GRB afterglow light curves (e.g., Dai \& Lu
1998a,b; Zhang \& M\'esz\'aros 2001). This is supported by the observations of both long and short GRBs
with the {\em Swift} mission (Fan \& Xu 2006; Liang et al. 2007; Troja et al. 2007; Lyons et al. 2010;
Rowlinson et al. 2010; Dall'Osso et al. 2011; Rowlinson et al. 2013; Gompertz et al. 2013, 2014; L\"{u}
\& Zhang 2014; L\"{u} et al. 2015; Gao et al. 2016; Gibson et al. 2017, 2018). A canonical X-ray light
curve was observed with the X-ray Telescope (XRT) on board the {\em Swift} mission for about half of
well-sampled X-ray afterglow light curves of the {\em Swift} long GRB sample. The canonical XRT
lightcurve in long GRBs is characterized with a shallow-decay segment, which transits to a normal-decay
segment or occasionally a sharp drop (Noseck et al. 2006; Zhang et al. 2006; O'Brien et al. 2006; Liang
et al. 2007; Troja et al. 2007; L\"{u} \& Zhang 2014; Du et al. 2016). The end time of this segment,
which may correspond to the characteristic spin-down timescale of the magnetar, ranges from tens of
seconds to several thousands or even to one day (Liang et al. 2007). Therefore, GRBs and their
afterglows may be probes to investigate the properties of newly-born magnetars (Yu et al. 2010; Metzger
et al. 2011; Yu et al. 2015).

The gravitational wave named as GW 170817 was discovered by advanced Laser Interferometer
Gravitational-wave Observatory (aLIGO)/Virgo detectors from double neutron stars (Abbott et al. 2016;
Abbott et al. 2017). Whether the survived magnetar can be a candidate as the remnant of those two
neutron stars merger remain debate (Abbott et al. 2017; Troja et al. 2017,2018; Margalit \& Metzger
2017; Drago \& Pagliara 2018; Ai et al. 2018). Here, we assume that the magnetar remnant can be
survived, a newly born magnetar should coherently lost their spin energy via both the magnetic dipole
radiation and gravitational wave radiation (Zhang \& M\'esz\'aros 2001; Giacomazzo \& Perna 2013; Fan
et al. 2013; Lasky et al. 2014; Lasky \& Glampedakis 2016). Inspired by this motivation, the similar
situation should be existence with magnerar central engine that is formed via massive star collapse
(Zhang \& M\'esz\'aros 2001; Yu et al. 2010; Yu et al. 2015), and it is supported by observed evidence
(Mazzali et al. 2014; L\"{u} et al. 2018). This paper theoretically presents the evolutional features
of the EM emission in the GW emission dominated and MD emission dominated scenarios when the magnetar
as the central engine of long GRBs. Then, we search for the corresponding observational evidence in the
X-ray afterglows of {\em Swift} long GRBs, and constrain the physical parameters of magnetar.
Throughout the paper, a concordance cosmology with parameters $H_0 = 71$ km s$^{-1}$ Mpc $^{-1}$,
$\Omega_M=0.30$, and $\Omega_{\Lambda}=0.70$ is adopted to calculate the luminosity of X-ray plateau.

\section{Magnetar Spin-Down by GW and EM Radiations}
The energy reservoir of a newly-born magnetar is its total rotation energy, which reads as
\begin{eqnarray}
E_{\rm rot} = \frac{1}{2} I \Omega^{2}
\simeq 2 \times 10^{52}~{\rm erg}~
M_{1.4} R_6^2 P_{-3}^{-2},
\label{Erot}
\end{eqnarray}
where $I$ is the moment of inertia, $\Omega$, $P$, $R$, and $M$ are the angular frequency, rotating
period, radius, and mass of the neutron star. The convention $Q = 10^x Q_x$ is adopted in cgs units. It
may spin down by losing its rotational energy through two channels, magnetic dipole torques ($L_{\rm
EM}$) and gravitational wave radiation ($L_{\rm GW}$) (e.g., Shapiro \& Teukolsky 1983; Zhang \&
M{\'e}sz{\'a}ros 2001; Fan et al. 2013; Lasky \& Glampedakis 2016), i.e.,
\begin{eqnarray}
-\frac{dE_{\rm rot}}{dt} = -I\Omega \dot{\Omega} &=& L_{\rm EM} + L_{\rm GW} \nonumber \\
&=& \frac{B^2_{\rm p}R^{6}\Omega^{4}}{6c^{3}}+\frac{32GI^{2}\epsilon^{2}\Omega^{6}}{5c^{5}},
\label{Spindown}
\end{eqnarray}
where $\dot{\Omega}$ is time derivative of angular frequency, $B_p$ is the surface magnetic field at
the pole, and $\epsilon=2(I_{\rm xx}-I_{\rm yy})/(I_{\rm xx}+I_{\rm yy})$ is the ellipticity in terms
of the principal moments of inertia. One can find that for a magnetar with given $R$ and $I$, its
$L_{\rm EM}$ depends on $B$ and $\Omega$, and $L_{\rm GW}$ depends on $\epsilon$ and $\Omega$. We
derive the evolution of $L_{\rm EM}$ in the phases of that $L_{\rm EM}$ and $L_{\rm GW}$ dominates the
rational energy lost in the following.

(I) $L_{\rm EM}$ dominated scenario:

In this scenario, one has
\begin{eqnarray}
L_{\rm EM}\simeq -I\Omega \dot{\Omega}=\frac{B^2_{\rm p}R^{6}\Omega^{4}}{6c^{3}}.
\label{EM_dominated}
\end{eqnarray}
The full solution of $\Omega(t)$ in Eq.(\ref{EM_dominated}) can be written as
\begin{eqnarray}
\Omega(t) &=& \Omega_{0}(1+\frac{t}{\tau_{\rm em}})^{-1/2} \nonumber \\
&\simeq&\cases{ \Omega_0,
            & t $\ll \tau_{\rm c,em}$  \cr
\Omega_{0}(\frac{t}{\tau_{\rm c,em}})^{-1/2},
            & t $\gg \tau_{\rm c,em}$ \cr
            }
\label{Omega_EM}
\end{eqnarray}
where $\Omega_{0}$ is initial angular frequency at $t=0$, and $\tau_{\rm c,em}$ is a characteristic
spin-down time scale in this scenario. $\tau_{\rm em}$  can be given by
\begin{eqnarray}
\tau_{\rm c,em}&=&\frac{3c^{3}I}{B_{p}^{2}R^{6}\Omega_{0}^{2}} \nonumber \\
&\simeq&2.05 \times 10^3~{\rm s}~ (I_{45} B_{p,15}^{-2} P_{0,-3}^2 R_6^{-6}),
\label{spintau_em}
\end{eqnarray}
where $P_0$ is the initial period of the magnetar (e.g., $P_0=2\pi/\Omega_0$). The evolution of $L_{\rm
EM}$ with time can be expressed as
\begin{eqnarray}
L_{\rm EM}(t) &=& L_{\rm em,0}(1+\frac{t}{\tau_{\rm c,em}})^{-2} \nonumber \\
&\simeq&\cases{L_{\rm em,0},
            & t $\ll \tau_{\rm c,em}$  \cr
L_{\rm em,0}(\frac{t}{\tau_{\rm c,em}})^{-2},
            & t $\gg \tau_{\rm c,em}$ \cr
            }
\label{Luminosity_EM}
\end{eqnarray}
where $L_{\rm em,0}$ is the initial kinetic luminosity of electromagnetic dipole emission at $t_0$,
given by
\begin{eqnarray}
L_{\rm em,0}&=&\frac{B^2_{p}R^6\Omega^{4}_{0}}{6c^3} \nonumber \\
&\simeq&1.0 \times 10^{49}~{\rm erg~s^{-1}} (B_{p,15}^2 P_{0,-3}^{-4} R_6^6),
\label{spinlu_em}
\end{eqnarray}

(II) $L_{\rm GW}$ dominated scenario:

In this scenario, one has
\begin{eqnarray}
L_{\rm GW}\simeq -I\Omega \dot{\Omega}=\frac{32GI^{2}\epsilon^{2}\Omega^{6}}{5c^{5}}.
\label{GW_dominated}
\end{eqnarray}
The full solution of $\Omega(t)$ in Eq.(\ref{GW_dominated}) can be written as
\begin{eqnarray}
\Omega(t) &=& \Omega_{0}(1+\frac{t}{\tau_{\rm c,gw}})^{-1/4} \nonumber \\
&\simeq&\cases{ \Omega_0,
            & t $\ll \tau_{\rm c,gw}$  \cr
\Omega_{0}(\frac{t}{\tau_{\rm c,gw}})^{-1/4},
            & t $\gg \tau_{\rm c,gw}$ \cr
            }
\label{Omega_GW}
\end{eqnarray}
where $\tau_{\rm c,gw}$ is a characteristic spin-down time scale in this scenario, which reads as
\begin{eqnarray}
\tau_{\rm c,gw}&=&\frac{5c^{5}}{128GI\epsilon^2\Omega^4_0} \nonumber \\
&\simeq&9.1 \times 10^3~{\rm s}~ (I^{-1}_{45}\epsilon_{-3}^{-2} P_{0,-3}^4 ).
\label{spintau_gw}
\end{eqnarray}
The evolution of $L_{\rm GW}$ with time thus can be expressed as
\begin{eqnarray}
L_{\rm GW}(t) &=& L_{\rm gw,0}(1+\frac{t}{\tau_{\rm c,gw}})^{-3/2} \nonumber \\
&\simeq&\cases{L_{\rm gw,0},
            & t $\ll \tau_{\rm c,gw}$  \cr
L_{\rm gw,0}(\frac{t}{\tau_{\rm c,gw}})^{-3/2},
            & t $\gg \tau_{\rm c,gw}$ \cr
            }
\label{Luminosity_GW}
\end{eqnarray}
where $L_{\rm GW,0}$ is the luminosity of gravitational-wave quadrupole emission at $t=0$, given by
\begin{eqnarray}
L_{\rm gw,0}&=& \frac{32GI^{2}\epsilon^{2}\Omega_0^{6}}{5c^{5}} \nonumber \\
&\simeq&1.08 \times 10^{48}~{\rm erg~s^{-1}}(I_{45}^2 \epsilon_{-3}^{2} P_{0,-3}^{-6}).
\label{spinlu_gw}
\end{eqnarray}
Within this scenario, the evolution of $\Omega(t)$ is different from the $L_{\rm EM}$ dominated
scenario, and the evolution of $L_{\rm EM}$ is replaced by
\begin{eqnarray}
L_{\rm EM}(t) &=& L_{\rm em,0}(1+\frac{t}{\tau_{\rm c,gw}})^{-1} \nonumber \\
&\simeq&\cases{L_{\rm em,0},
            & t $\ll \tau_{\rm c,gw}$  \cr
L_{\rm em,0}(\frac{t}{\tau_{\rm c,gw}})^{-1},
            & t $\gg \tau_{\rm c,gw}$ \cr
            }
\label{Luminosity_GWEM}
\end{eqnarray}

Based on Eq.(\ref{Luminosity_EM}) and Eq.(\ref{Luminosity_GWEM}), and following the method of Lasky \&
Glampedakis (2016), one can obtain the transition time ($\tau_{\ast}$) which point is from
gravitational-wave quadrupole dominated to electromagnetic dipole dominated emission (Zhang \&
M{\'e}sz{\'a}ros 2001; Lasky \& Glampedakis 2016),
\begin{eqnarray}
\tau_{\ast}=\frac{\tau_{\rm c,em}}{\tau_{\rm c,gw}}(\tau_{\rm c,em}-2\tau_{\rm c,gw})
\label{transition_time}
\end{eqnarray}
One can observe $\tau_{\ast}<0$ if $\tau_{\rm c,em}<2\tau_{\rm c,gw}$, indicating that no time is found
for making $L_{\rm EM}(t)=L_{\rm GW}(t)$ and the rotational energy lost is always dominated by
electromagnetic dipole emission, and the $L_{EM}$ evolves with time as Eq. (\ref{Luminosity_EM}). If
$\tau_{\rm c,em}>2\tau_{\rm c,gw}$, one has $\tau_{\ast}>0$, suggesting that the spin-down is dominated
by gravitational-wave quadrupole early and electromagnetic dipole emission later. In this case, the
luminosity evolves with time as
\begin{eqnarray}
L(t)&\propto&\cases{(1+\frac{t}{\tau_{\rm c,gw}})^{-1},
            & t $\leq\tau_{\ast}$ (GW dominated)  \cr
t^{-2},
            &t $>\tau_{\ast}$ (EM dominated). \cr
            }
\label{L_EMGW}
\end{eqnarray}

Moreover, Usov (1992) proposed that the spin-down of the neutron star is dominated by
gravitational-wave quadrupole radiation when the angular velocity $\Omega$ of magnetar is larger than
the critical value $\Omega_{cr}$ (also see Blackman \& Yi 1998; Zhang \& M{\'e}sz{\'a}ros 2001). Here,
the $\Omega_{cr}$ ranges in $(0.4\sim 1.2)\times 10^{4}\rm~s^{-1}$ that depended on the equation of
state of neutron star.

\section{A New Paradigm for Physical Origin of the X-ray Afterglows}
\begin{figure*}
\includegraphics[angle=0,scale=0.3]{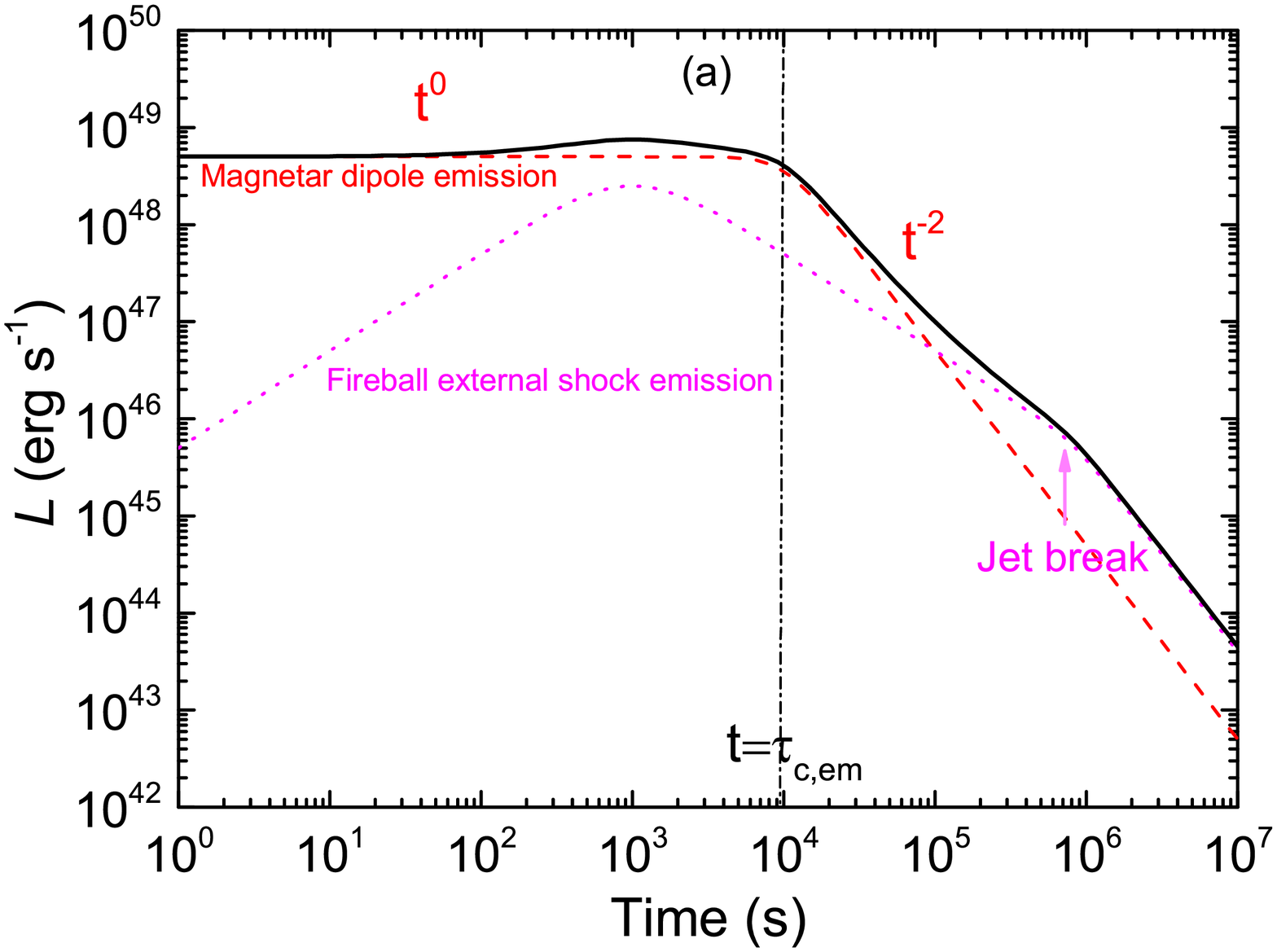}
\includegraphics[angle=0,scale=0.3]{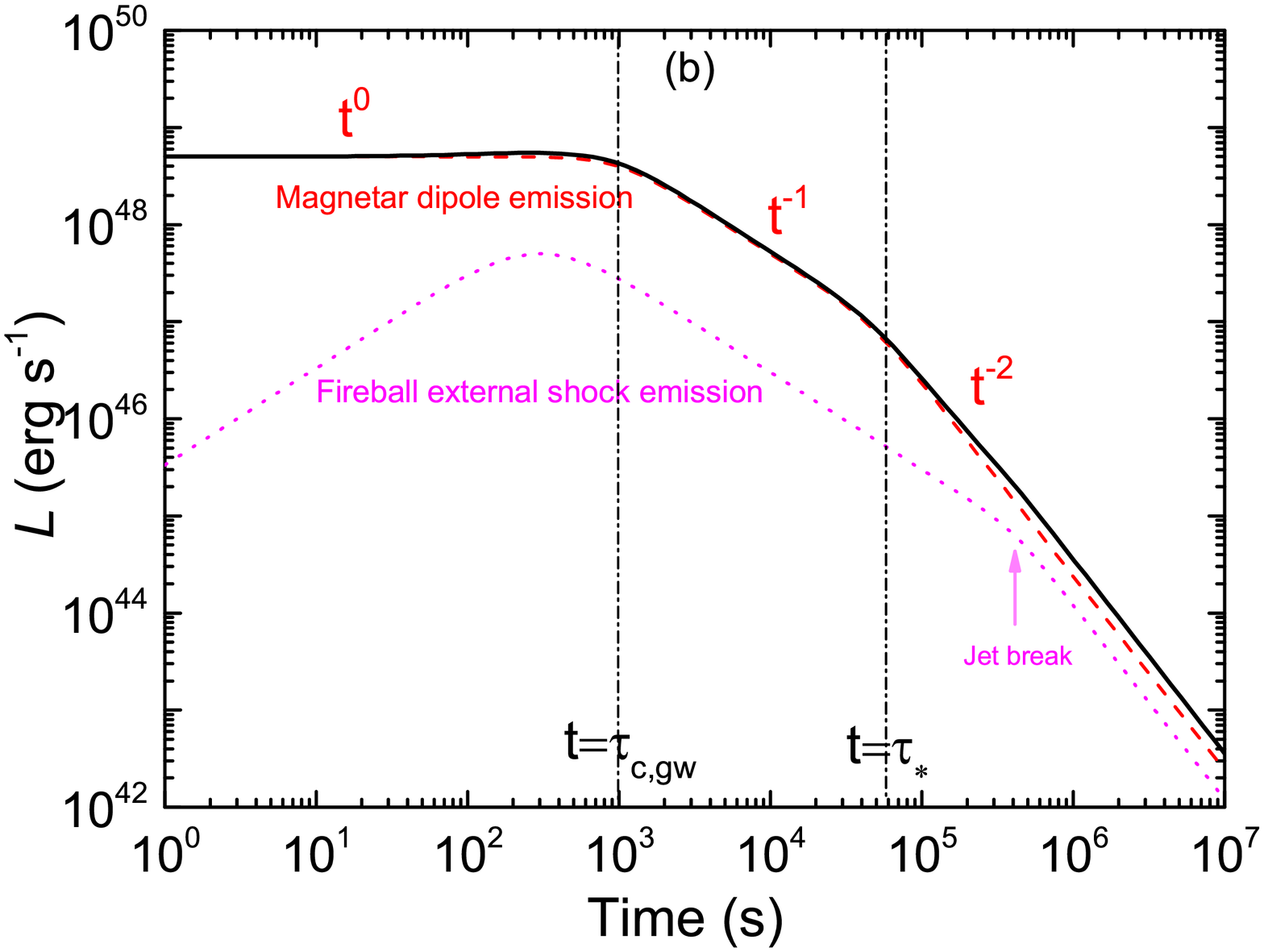}
\includegraphics[angle=0,scale=0.3]{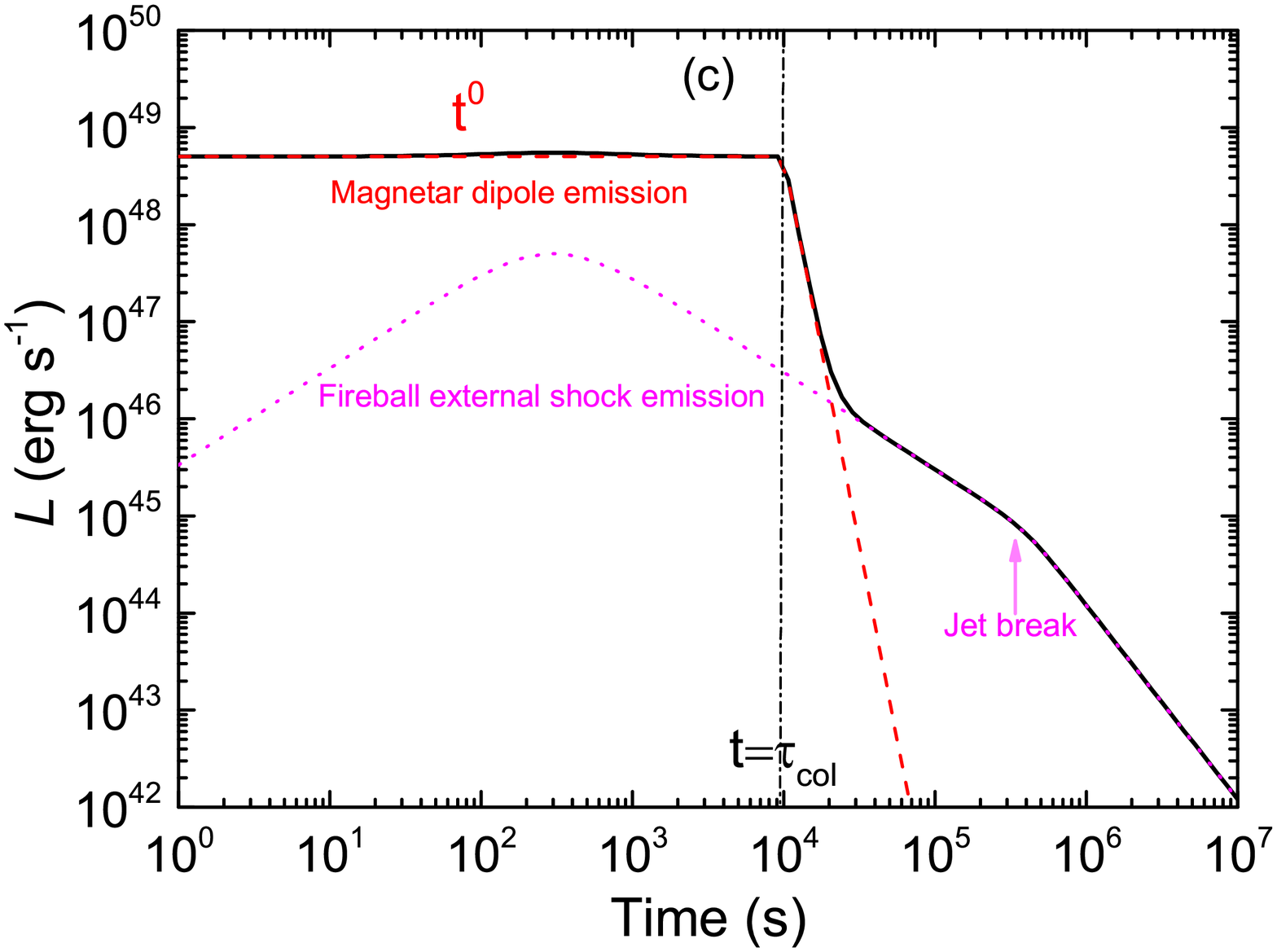}
\hfill\center \caption{Illustrations of the afterglow lightcurves of GRBs with a newly-born magnetar as
their central engines in different scenarios. (a):  A stable magnetar spins down via magnetic dipole
radiation. (b): A stable magnetar spins down via gravitational-wave quadrupole emission early
($t<\tau_{\ast}$) and magnetic dipole radiation later ($t>\tau_{\ast}$). (c): A supra magnetar
collapses to form a black hole. The red-dashed lines represent the emission from the magnetic dipole
radiation, the pink-dotted lines stand for the afterglows of the GRB fireballs as predicted by the
standard external shock model in the thin shell case (e.g., M\'esz\'aros \& Rees 1997; Sari et al.
1998). The black-solid lines are the total lightcurves contributed by both the dipole wind and external
shocks of the GRB fireball.} \label{fig:cartoon}
\end{figure*}

Based on our derivations above, a light curve of magnetic dipole emission of a newly-born magnetar may
be characterized with some segments, as illustrate in Fig.\ref{fig:cartoon}. It starts with a segment
of $F\propto t^{0}$ for both the magnetic dipole emission and GW emission dominated scenarios. It
transits to $L\propto t^{-2}$ at $t_{\rm c,em}$ if the magnetar still stable and its spin-down is
always dominated by the magnetic dipole radiation, as that shown in Fig.\ref{fig:cartoon} (a). If the
spin-down of the magnetar is initially dominated by the GW emission, the $L_{\rm em}$ light curve
should transits $L\propto t^{-1}$ at $t_{\rm c,gw}$, and decays as $L\propto t^{-2}$ since
$t=\tau_{\ast}$, at then the magnetic dipole radiation begins dominating the rotation energy lost, as
shown in Fig.\ref{fig:cartoon} (b). Occasionally, the magnetar may be collapsed to a black hole before
$t_{\rm c,gw}$ and $t_{\rm c,em}$, the magnetic dipole radiation is shut down suddenly, the light curve
shows up as an internal plateau followed by a sharp drop, as shown in Fig.\ref{fig:cartoon} (c). Note
that the discussion above is in the two extremely cases, i.e., the magnetic dipole radiation and GW
emission dominated cases. In a generic case, the magnetar coherently spins down by both the GW emission
and magnetic dipole radiation, the $L_{\rm em}$ light curve may be featured as a shallow-to-normal
decay segment with a decay slope changing from 0 to $-1\sim -2$. This resembles the canonical XRT light
curve of {\em Swift} GRBs presented by Zhang et al. (2006).

\begin{figure*}
\includegraphics[angle=0,scale=0.3]{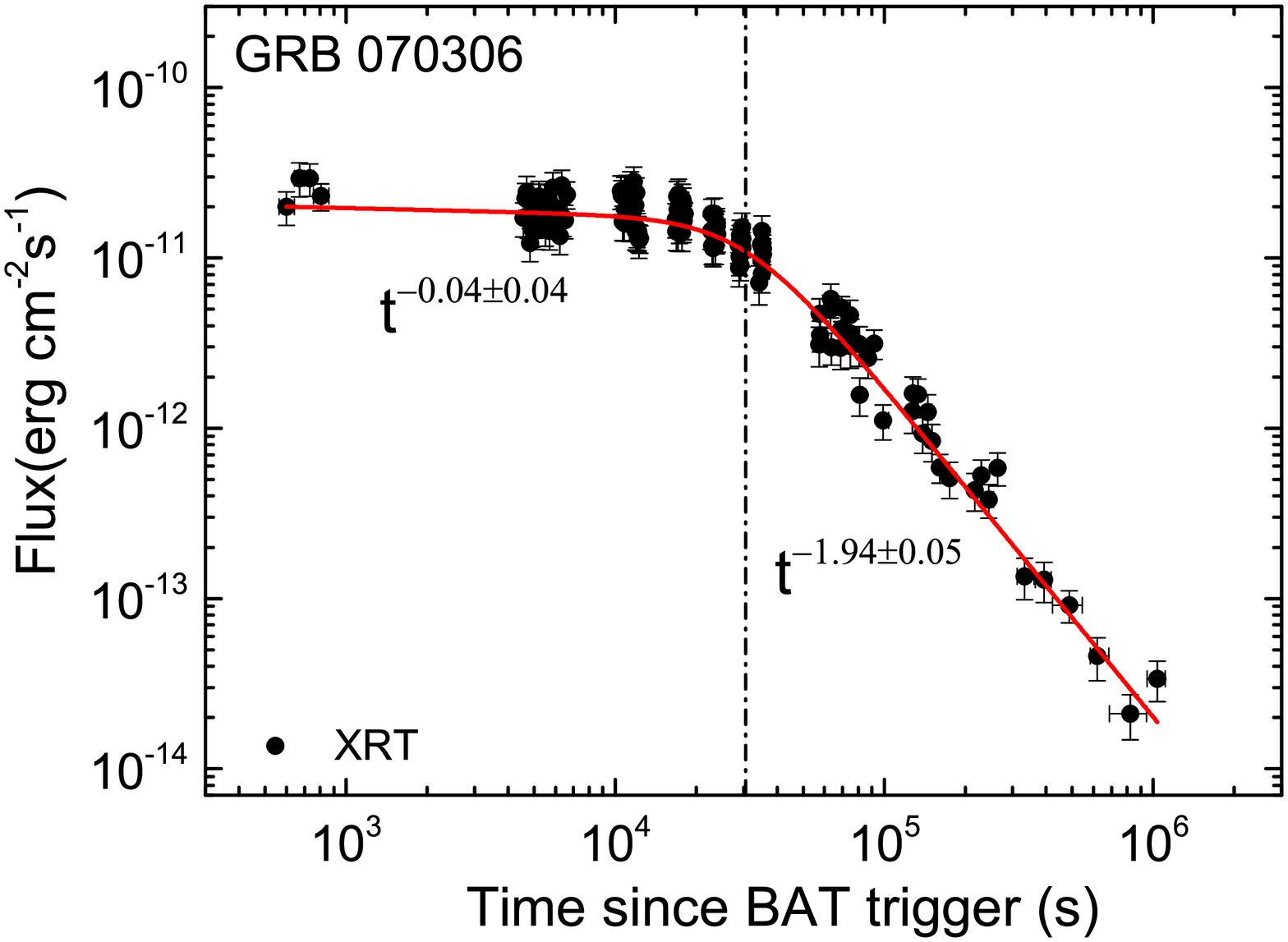}
\includegraphics[angle=0,scale=0.3]{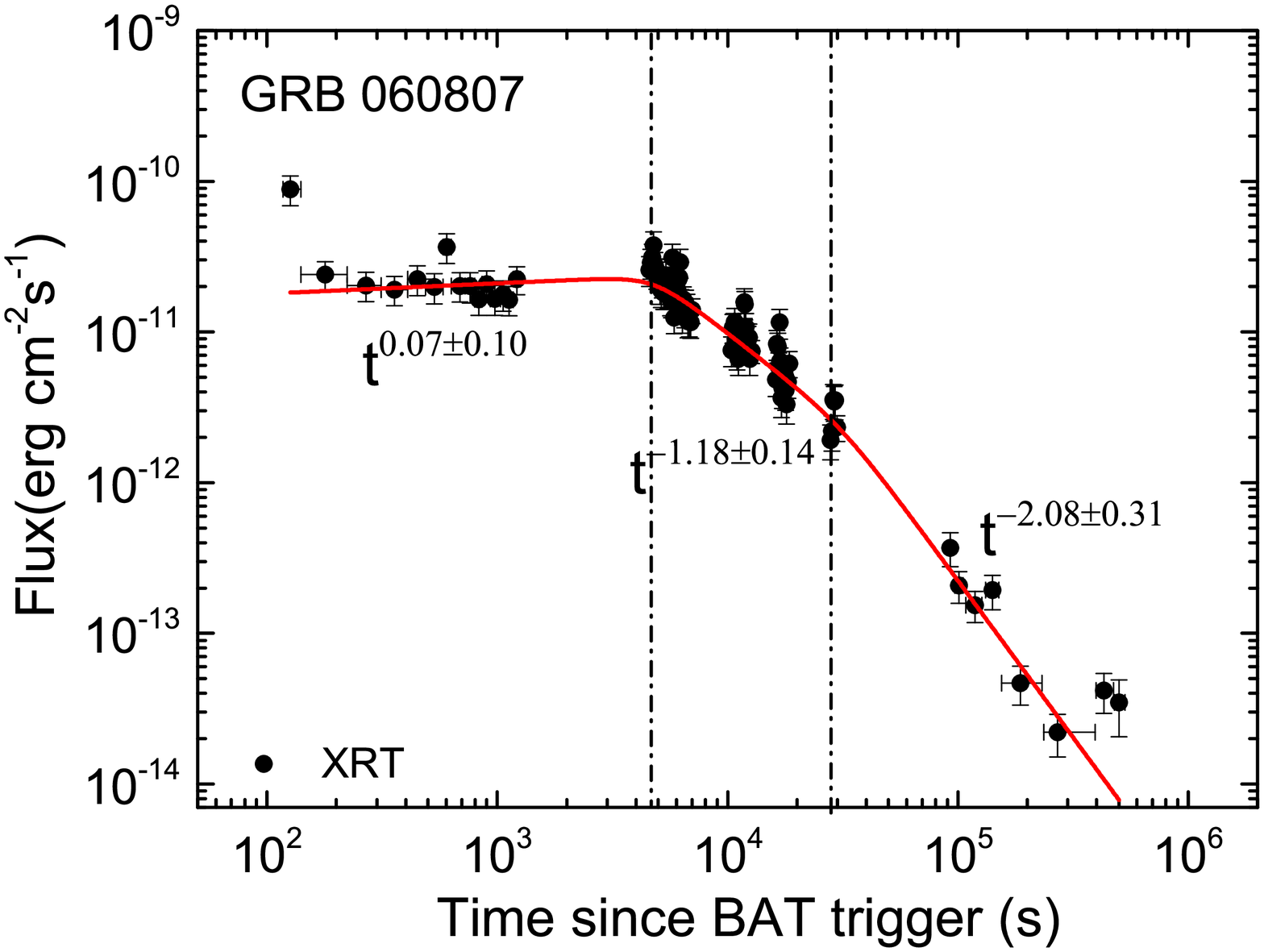}
\includegraphics[angle=0,scale=0.3]{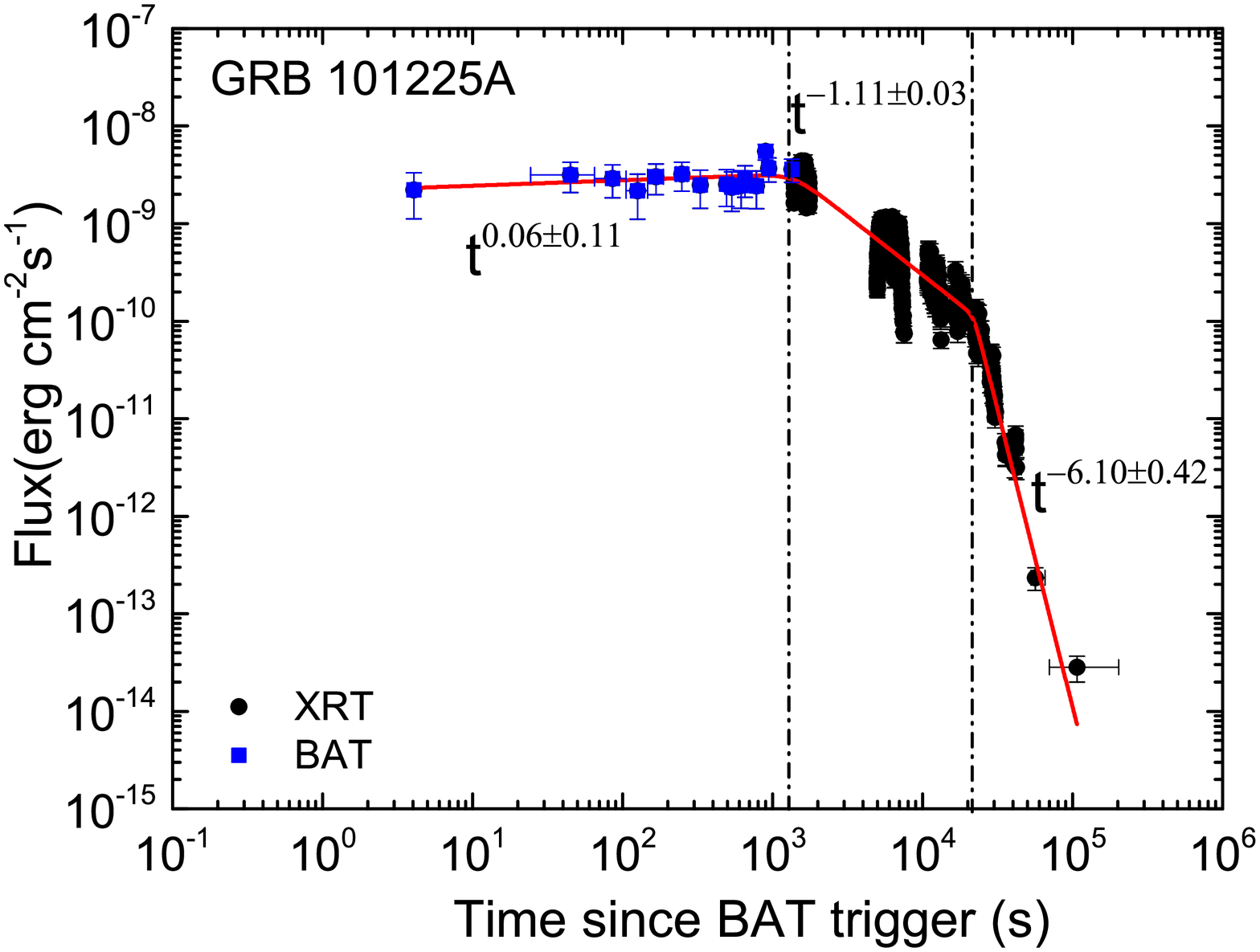}
\includegraphics[angle=0,scale=0.3]{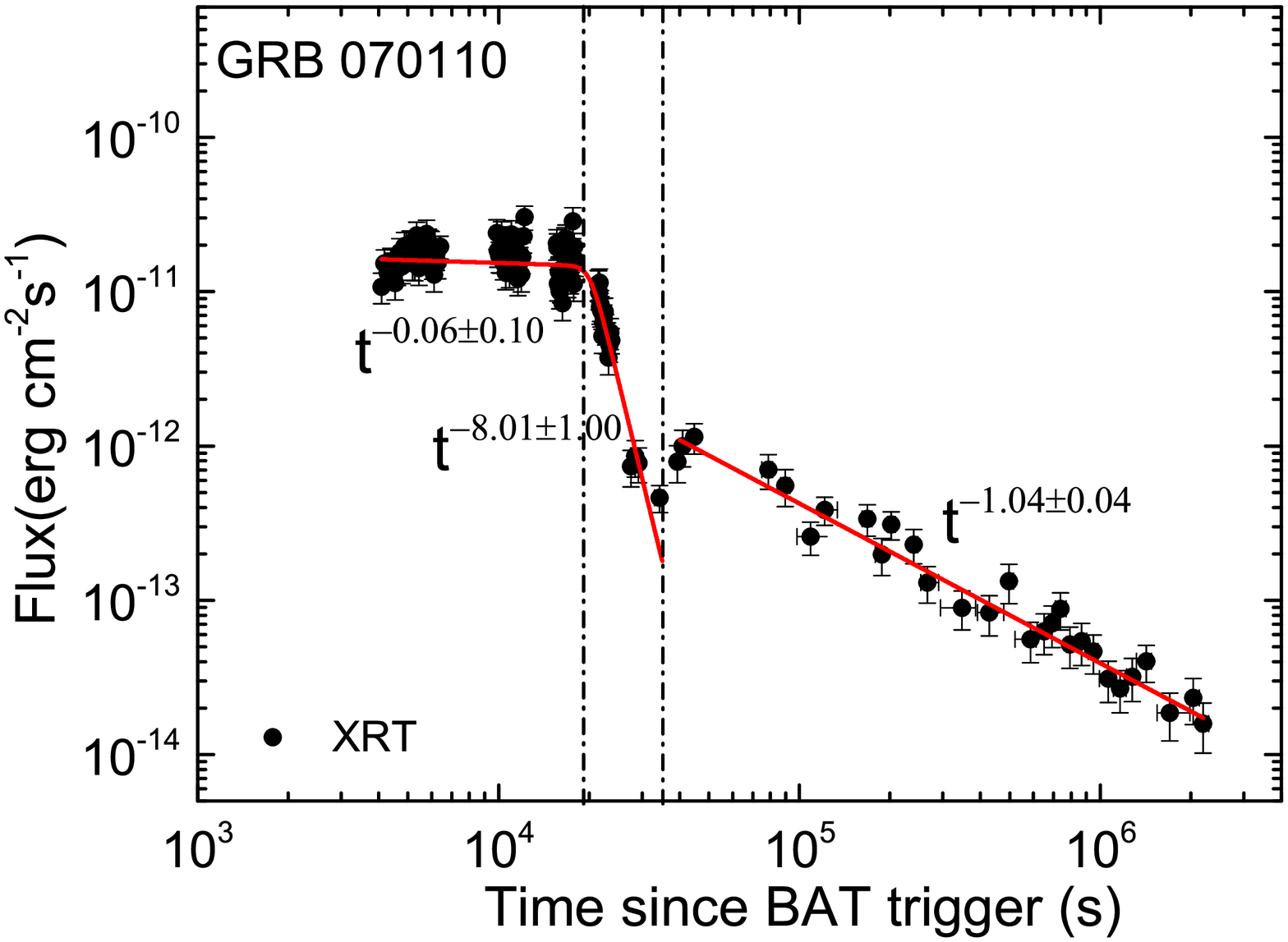}
\hfill\center \caption{Examples of X-ray light curves (dots) and our power-law fits (red lines) of four
GRBs that may be regarded as prototype light curves of dipole radiations in the scenarios shown in Fig.
1: GRB 070306---Scenario (a): a stable magnetar spinning down via magnetic dipole radiation; GRB 060807
---Scenario (b): a stable magnetar spinning down via gravitational-wave quadrupole emission early and
magnetic dipole radiation later; GRB 101225---Scenario (b) and (c): collapse of a supra magnetar during
its spinning down period via gravitational-wave quadrupole emission; GRB 070110---Scenario (c):
collapse of a supra magnetar before the spin-down characteristic timescale.} \label{fig:BATXRTLC}
\end{figure*}

The combination of the magnetic dipole radiation and GRB fireball afterglow emission may make the
diversity of the observed light curves in long GRBs. We also illustrate the afterglow light curve in
the thin shell case and the total light curve of both the magnetic dipole radiation and GRB afterglows
in Fig.\ref{fig:cartoon}. Zou et al. (2018) present a systematical analysis for the XRT light curves of
{\em Swift} GRBs in the past 13 years and got a sample of 111 long GRBs that their early XRT light
curves show a shallow-decay segment. As shown in their Figure 3, the XRT light curves of a large
fraction of long GRBs (94/111) have a shallow-to-normal-decaying segment. The slope of the
normal-decaying segment varies from $\sim -1$ to $\sim -2$, which extends up to hours even days (seel
also Liang et al. 2007), being consistent with the generic scenario of our paradigm. We try to find out
evidences of magnetar spin-down via gravitational wave quadrupole emission and magnetic dipole
radiation from observations, and derive the parameters of magnetar. Some XRT lightcurves of interest
long GRBs are presented in the following:

\begin{itemize}
\item GRBs 070306. As shown in Fig.\ref{fig:BATXRTLC}(a), its XRT lightcurve is a prototype
    lightcurve of the scenario that the magnetar is stable and the magnetic dipole emission may
    dominate the rotation energy lost. Its flux decay slope changes from $(-0.04\pm 0.04)$ to
    $(-1.94\pm 0.05)$ at $(3.06\pm 0.22)\times 10^{4}$ seconds. This is consistent with prediction
    of magnetar spin-down with magnetic dipole emission dominated. The redshift of GRB 070307 is
    1.496, the plateau luminosity and break time are rough equal to $L_{em,0}$ and $\tau_{c,em}$ in
    the rest frame, respectively. Based on the Eq.\ref{spintau_em} and Eq.\ref{spinlu_em}, one can
    constrain the $B_{\rm p}\sim(3.18\pm0.37)\times 10^{15}$ G, $P_0\sim (4.29\pm0.34)$ ms.
\item GRB 060807. As shown in Fig.\ref{fig:BATXRTLC}(b), a transition from the GW emission
    dominated epoch to the magnetic dipole radiation dominated epoch is detected in this GRB. Its
    XRT lightcurve is a prototype lightcurve of the scenario that the magnetar is stable and its
    rotation energy lost initially is dominated by the GW emission. Excluding the very early steep
    decay segment that may be attributed to the tail emission of the prompt gamma-rays (Liang et
    al. 2007), its XRT lightcurve can fit with a triple power-law function with reduced
    $\chi^2=1.18$. The slope changes from $\alpha_0=(-0.07\pm 0.09)$ to $\alpha_1=(1.18\pm 0.13)$
    at $t_1=(4.80\pm1.10)\times 10^{3}$ s, and finally $\alpha_2=(2.08\pm 0.13)$ at
    $t_2=(2.86\pm0.78)\times 10^4$ s\footnote{One broken power-law model is also invoked to fit the
    same data, one has the slopes with -0.06 and -1.76 before and after break time, respectively.
    However, the reduced $\chi^2$ is as large as 1.55. This is too large to be accepted from the
    statistical point of view.}, as shown in Fig.\ref{fig:BATXRTLC}. Moreover, the Swift XRT team's
    analysis also finds that two breaks are statistically necessary with an f-test revealing a
    $5\times 10^{4}\%$ probability of chance improvement (Evans et al 2007, 2009). The
    characteristic of this light curve is consistent with a stable magnetar spinning down via
    gravitational wave quadrupole emission early (before $t_2$) and magnetic dipole radiation later
    (after $t_2$). Also, those two break times may be the spin-down characteristic $\tau_c$ of the
    magnetar and the transition time $\tau_{\ast}$ from the GW emission dominated epoch to the
    magnetic dipole radiation dominated epoch, respectively. Based on the derivations in section 2,
    together with $\tau_{\rm c, gw}\sim t_1$ and $\tau_{\ast}\sim t_2$, one has $\tau_{\rm c,
    em}=(1.75\pm 0.46)\times10^4$ s. By assuming redshift $z=0.5$ of this case, one can constrain
    the physical parameters of magnetar with equation of state GM1, e.g., $B_{\rm
    p}\sim(7.53\pm3.43)\times 10^{14}$ G, $P_0\sim (1.16\pm0.56)$ ms ($\Omega_0\sim
    5400\rm~s^{-1}$), and $\epsilon \sim (1.02\pm0.35)\times10^{-3}$. It is consistent with the
    required of GW losses significantly with a new born neutron star\footnote{Here, we assume that
    the properties of neutron star is similar with that magnetar. So that, one can use similar
    method in Lasky \& Glampedakis 2016 to estimate the ellipticity of magnetar.}.
\item GRB 101225A is of interest with detection of a possible signal that the magnetar is collapsed
    to a black hole before transition from the GW emission dominated epoch to the magnetic dipole
    radiation dominated epoch. As shown in Fig.\ref{fig:BATXRTLC}(c), it XRT lightcurve is roughly
    depicted with a triple power-law function without considering the significant flickering. The
    decaying slope is initially $\alpha_0=(-0.05\pm 0.11)$, then transits to $\alpha_1=(-1.11\pm
    0.03)$ at $t_1=(1.42\pm 0.15)\times 10^3$ s, and finally becomes $\alpha_2=(-5.93\pm 0.31)$ at
    $t_2=(2.12\pm0.9)\times 10^4$ s. The sharp break at $t_2$ may be due to the collapse of the
    magnetar. Based on the derivations of section 2 and light curves fits above, one has $\tau_{\rm
    c, GW}\simeq t_1=(1.42\pm 0.15)\times 10^3$ s and $\tau_{\ast}> t_2=(2.12\pm0.9)\times 10^4$ s.
    Together with Eq.(\ref{transition_time}), one can estimate $\tau_{\rm em}>7077$ s. On the other
    hand, $L_{\rm em,0}\simeq 4\pi D_L^2 F_b=(6.15\pm0.54)\times 10^{48}~{\rm erg~s^{-1}}$ ($F_b$
    is flux at $t_1$ in X-ray light curve) at redshift $z=0.847$, one can estimate the $B_{\rm
    p}<5.8\times 10^{15}$ G, $P_0<1.35$ ms, and $\epsilon>1.7\times10^{-3}$ with equation of state
    GM1 (Lasky et al. 2014; Ravi \& Lasky 2014; L\"{u} et al. 2015).
\item GRB 070110. As shown in Fig.\ref{fig:BATXRTLC}(d), its XRT lightcurve is a plateau followed
    by a sharp drop (Troja et al. 2007). This may be a signature of collapse of a magnetar before
    the characteristic spin-down timescale. The collapse of the magnetar makes the magnetic dipole
    radiation disappears suddenly and the GRB afterglows emerge, as that observed since $t>6\times
    10^{4}$ seconds. Similar feature is also observed in GRBs 060602, 070616, 060607A, and 170714
    (Zou et al. 2018). .
\end{itemize}

\section{Conclusions and Discussion}
We have presented the temporal evolution feature of the EM emission during the spin-down of a
newly-born magnetar (the remnant of massive star collapse) in the scenarios that the GW quadrupole and
magnetic dipole emission dominate the rotational energy lost, respectively. We show that the EM
emission light curve of the magnetic dipole radiation is described as $F\propto (1+t/\tau_c)^{\alpha}$,
where $\tau_c$ is the characteristic spin-down timescales, and $\alpha=-1$ in the GW emission dominated
scenario and $\alpha=-2$ in the magnetic dipole radiation dominated scenario. Transition from a GW
emission dominated epoch to a magnetic dipole radiation dominated epoch may show up as a smooth break
with a decaying slope from $-1$ to $-2$. In case of that the magnetar collapses to a black hole before
$\tau_c$, the light curve of EM emission should be a plateau followed by a sharp drop being due to the
immediate shut down of the magnetic dipole radiation. In a generic scenario, the magnetar coherently
spins down by both the GW emission and magnetic dipole radiation, the EM emission lightcurve may be a
shallow-to-normal decay segment with a slope changing from 0 to $-1\sim -2$.

Our result presents a new paradigm for physical origin to explain the X-ray data of long GRBs observed
with XRT. The expected generic light curve in this framework resembles the canonical XRT light curve of
{\em Swift} GRBs presented by Zhang et al. (2006). Analysis of XRT data for a large sample of long GRBs
and cases study are also consistent with our paradigm (e.g., Zou et al. 2018). We propose that the
early shallow-decaying X-ray afterglows in long GRBs may be dominated by the magnetic dipole emission
but not external shock mission of GRB fireballs. The combination of the magnetic dipole radiation and
GRB fireball afterglows may make the diversity of the observed X-ray light curves. As shown in Liang et
al. (2010, 2013) and Li et al. (2012), a clear onset bump, which may be a signature of the deceleration
of GRB fireballs, is observed in the optical afterglows of some GRBs. Their XRT light curves in the
according time interval are usually a shallow-decay segment. It is possible that the X-ray emission is
dominated by the magnetic dipole radiation and the optical emission is dominated by the GRB afterglows.
This may interpret the observed chromatic behaviors of the X-ray and optical afterglow light curves and
the extremely low detection rate of a jet-like break in the X-ray light curves (Uhm \& Beloborodov
2007; Genet et al. 2007; Liang et al. 2008).

Moreover, we present four long GRBs (070306, 060807, 070110, and 101225A) which X-ray light curves are
likely contributed from magnetar spin-down via gravitational wave quadrupole emission and magnetic
dipole radiation. Such as GRB 070306, the X-ray emission is likely originated from magnetar spin-down
with magnetic dipole emission dominated; or magnetar spin-down via gravitational wave quadrupole
emission and magnetic dipole radiation (e.g., GRB 060807); or magnetar collapse into black hole before
its spin down (GRB 070110); or collapse of a supra magnetar during its spinning down period via
gravitational wave quadrupole emission (GRB 101225A). Within this scenario, the physical parameters of
magnetar can also be constrained via the observed properties of X-ray emission, e.g., $B_p$, $P_0$ and
$\epsilon$. Especially, the derived value of $\epsilon$ for long GRBs 060807 and 101225A are consistent
with the constraints of Lasky \& Glampedakis 2016 by adopting short GRBs.

On the other hand, the origin of some peculiar (e.g., soft-long lasting $\gamma$-ray emission)
Ultra-long GRBs may be consistent with our scenario. For example, the possible origin of GRB 101225A
may be consistent with the hypothesis magnetar spin-down with GW emission dominated and collapses to a
black hole at later time, e.g., long-soft $\gamma$-ray emission may be of energy injection from
magnetar wind within off-axis observations, the early segment of normal decay in XRT light curve is
consistent with spin-down magnetar wind with gravitational-wave quadrupole radiation dominated, and the
later phase of steeper decay in XRT light curve is the signature of magnetar collapse into black hole.
However, due to lack of smoking gun evidence, catching this possible interpretation is expected by
observing more events of GW and EM associated with LIGO and high energy instruments in the future.

\section{Acknowledgements}
We acknowledge the use of the public data from the {\em Swift} data archive and the UK {\em Swift}
Science Data Center, and thank the anonymous referee for helpful comments. This work is supported by
the National Basic Research Program (973 Programme) of China 2014CB845800, the National Natural Science
Foundation of China (Grant No.11603006, 11851304, 11533003 and U1731239), Guangxi Science Foundation
(grant No. 2017GXNSFFA198008, 2016GXNSFCB380005 and AD17129006), the One-Hundred-Talents Program of
Guangxi colleges, the high level innovation team and outstanding scholar program in Guangxi colleges,
Scientific Research Foundation of Guangxi University (grant no XGZ150299), and special funding for
Guangxi distinguished professors (Bagui Yingcai \& Bagui Xuezhe).



\begin{thebibliography}{99}
\bibitem[\protect\citeauthoryear{Abbott et al.}{2016}]{2016PhRvL.116f1102A} Abbott B.~P., et al., 2016,
    PhRvL, 116, 061102


\bibitem[\protect\citeauthoryear{Abbott et al.}{2017}]{2017PhRvL.119p1101A} Abbott B.~P., et al., 2017,
    PhRvL, 119, 161101


\bibitem[\protect\citeauthoryear{Ai et al.}{2018}]{2018ApJ...860...57A} Ai S., Gao H., Dai Z.-G., Wu
    X.-F., Li A., Zhang B., Li M.-Z., 2018, ApJ, 860, 57


\bibitem[\protect\citeauthoryear{Blackman \& Yi}{1998}]{1998ApJ...498L..31B} Blackman E.~G., Yi I.,
    1998, ApJ, 498, L31



\bibitem[\protect\citeauthoryear{Bucciantini et al.}{2012}]{2012MNRAS.419.1537B} Bucciantini N.,
    Metzger B.~D., Thompson T.~A., Quataert E., 2012, MNRAS, 419, 1537


\bibitem[\protect\citeauthoryear{Dai \& Lu}{1998}]{1998PhRvL..81.4301D} Dai Z.~G., Lu T., 1998, PhRvL,
    81, 4301


\bibitem[\protect\citeauthoryear{Dai \& Lu}{1998}]{1998A&A...333L..87D} Dai Z.~G., Lu T., 1998, A\&A,
    333, L87


\bibitem[\protect\citeauthoryear{Dall'Osso et al.}{2011}]{2011A&A...526A.121D} Dall'Osso S., Stratta
    G., Guetta D., Covino S., De Cesare G., Stella L., 2011, A\&A, 526, A121



\bibitem[\protect\citeauthoryear{Drago \& Pagliara}{2018}]{2018ApJ...852L..32D} Drago A., Pagliara G.,
    2018, ApJ, 852, L32


\bibitem[\protect\citeauthoryear{Du et al.}{2016}]{2016MNRAS.462.2990D} Du S., L{\"u} H.-J., Zhong
    S.-Q., Liang E.-W., 2016, MNRAS, 462, 2990

\bibitem[\protect\citeauthoryear{Evans et al.}{2009}]{2009MNRAS.397.1177E} Evans P.~A., et al.,
2009, MNRAS, 397, 1177


\bibitem[\protect\citeauthoryear{Evans et al.}{2007}]{2007A&A...469..379E} Evans P.~A., et al.,
2007, A\&A, 469, 379


\bibitem[\protect\citeauthoryear{Fan, Wu, \& Wei}{2013}]{2013PhRvD..88f7304F} Fan Y.-Z., Wu X.-F., Wei
    D.-M., 2013, PhRvD, 88, 067304


\bibitem[\protect\citeauthoryear{Fan \& Xu}{2006}]{2006MNRAS.372L..19F} Fan Y.-Z., Xu D., 2006, MNRAS,
    372, L19


\bibitem[\protect\citeauthoryear{Gao, Zhang, \& L{\"u}}{2016}]{2016PhRvD..93d4065G} Gao H., Zhang B.,
    L{\"u} H.-J., 2016, PhRvD, 93, 044065


\bibitem[\protect\citeauthoryear{Genet, Daigne, \& Mochkovitch}{2007}]{2007MNRAS.381..732G} Genet F.,
    Daigne F., Mochkovitch R., 2007, MNRAS, 381, 732


\bibitem[\protect\citeauthoryear{Giacomazzo \& Perna}{2013}]{2013ApJ...771L..26G} Giacomazzo B., Perna
    R., 2013, ApJ, 771, L26



\bibitem[\protect\citeauthoryear{Gibson et al.}{2018}]{2018MNRAS.478.4323G} Gibson S.~L., Wynn G.~A.,
    Gompertz B.~P., O'Brien P.~T., 2018, MNRAS, 478, 4323


\bibitem[\protect\citeauthoryear{Gibson et al.}{2017}]{2017MNRAS.470.4925G} Gibson S.~L., Wynn G.~A.,
    Gompertz B.~P., O'Brien P.~T., 2017, MNRAS, 470, 4925


\bibitem[\protect\citeauthoryear{Gompertz, O'Brien, \& Wynn}{2014}]{2014MNRAS.438..240G} Gompertz
    B.~P., O'Brien P.~T., Wynn G.~A., 2014, MNRAS, 438, 240


\bibitem[\protect\citeauthoryear{Gompertz et al.}{2013}]{2013MNRAS.431.1745G} Gompertz B.~P., O'Brien
    P.~T., Wynn G.~A., Rowlinson A., 2013, MNRAS, 431, 1745


\bibitem[\protect\citeauthoryear{Kumar \& Zhang}{2015}]{2015PhR...561....1K} Kumar P., Zhang B., 2015,
    PhR, 561, 1


\bibitem[\protect\citeauthoryear{L{\"u} et al.}{2018}]{2018arXiv180606249L} L{\"u} H.-J., Lan L., Zhang
    B., Liang E.-W., Kann D.~A., Du S.-S., Shen J., 2018, arXiv, arXiv:1806.06249



\bibitem[\protect\citeauthoryear{L{\"u} \& Zhang}{2014}]{2014ApJ...785...74L} L{\"u} H.-J., Zhang B.,
    2014, ApJ, 785, 74


\bibitem[\protect\citeauthoryear{L{\"u} et al.}{2015}]{2015ApJ...805...89L} L{\"u} H.-J., Zhang B., Lei
    W.-H., Li Y., Lasky P.~D., 2015, ApJ, 805, 89


\bibitem[\protect\citeauthoryear{Lasky \& Glampedakis}{2016}]{2016MNRAS.458.1660L} Lasky P.~D.,
    Glampedakis K., 2016, MNRAS, 458, 1660


\bibitem[\protect\citeauthoryear{Lasky et al.}{2014}]{2014PhRvD..89d7302L} Lasky P.~D., Haskell B.,
    Ravi V., Howell E.~J., Coward D.~M., 2014, PhRvD, 89, 047302


\bibitem[\protect\citeauthoryear{Lei et al.}{2009}]{2009ApJ...700.1970L} Lei W.~H., Wang D.~X., Zhang
    L., Gan Z.~M., Zou Y.~C., Xie Y., 2009, ApJ, 700, 1970


\bibitem[\protect\citeauthoryear{Li et al.}{2012}]{2012ApJ...758...27L} Li L., et al., 2012, ApJ, 758,
    27


\bibitem[\protect\citeauthoryear{Liang et al.}{2013}]{2013ApJ...774...13L} Liang E.-W., et al., 2013,
    ApJ, 774, 13


\bibitem[\protect\citeauthoryear{Liang et al.}{2008}]{2008ApJ...675..528L} Liang E.-W., Racusin J.~L.,
    Zhang B., Zhang B.-B., Burrows D.~N., 2008, ApJ, 675, 528


\bibitem[\protect\citeauthoryear{Liang et al.}{2010}]{2010ApJ...725.2209L} Liang E.-W., Yi S.-X., Zhang
    J., L{\"u} H.-J., Zhang B.-B., Zhang B., 2010, ApJ, 725, 2209


\bibitem[\protect\citeauthoryear{Liang, Zhang, \& Zhang}{2007}]{2007ApJ...670..565L} Liang E.-W., Zhang
    B.-B., Zhang B., 2007, ApJ, 670, 565


\bibitem[\protect\citeauthoryear{Liu, Gu, \& Zhang}{2017}]{2017NewAR..79....1L} Liu T., Gu W.-M., Zhang
    B., 2017, NewAR, 79, 1


\bibitem[\protect\citeauthoryear{Lyons et al.}{2010}]{2010MNRAS.402..705L} Lyons N., O'Brien P.~T.,
    Zhang B., Willingale R., Troja E., Starling R.~L.~C., 2010, MNRAS, 402, 705


\bibitem[\protect\citeauthoryear{M{\'e}sz{\'a}ros \& Rees}{1997}]{1997ApJ...476..232M} M{\'e}sz{\'a}ros
    P., Rees M.~J., 1997, ApJ, 476, 232


\bibitem[\protect\citeauthoryear{Margalit \& Metzger}{2017}]{2017ApJ...850L..19M} Margalit B., Metzger
    B.~D., 2017, ApJ, 850, L19


\bibitem[\protect\citeauthoryear{Mazzali et al.}{2014}]{2014MNRAS.443...67M} Mazzali P.~A., McFadyen
    A.~I., Woosley S.~E., Pian E., Tanaka M., 2014, MNRAS, 443, 67



\bibitem[\protect\citeauthoryear{Metzger et al.}{2011}]{2011MNRAS.413.2031M} Metzger B.~D., Giannios
    D., Thompson T.~A., Bucciantini N., Quataert E., 2011, MNRAS, 413, 2031


\bibitem[\protect\citeauthoryear{Metzger \& Piro}{2014}]{2014MNRAS.439.3916M} Metzger B.~D., Piro
    A.~L., 2014, MNRAS, 439, 3916


\bibitem[\protect\citeauthoryear{Narayan, Piran, \& Kumar}{2001}]{2001ApJ...557..949N} Narayan R.,
    Piran T., Kumar P., 2001, ApJ, 557, 949


\bibitem[\protect\citeauthoryear{Nousek et al.}{2006}]{2006ApJ...642..389N} Nousek J.~A., et al., 2006,
    ApJ, 642, 389


\bibitem[\protect\citeauthoryear{O'Brien et al.}{2006}]{2006ApJ...647.1213O} O'Brien P.~T., et al.,
    2006, ApJ, 647, 1213


\bibitem[\protect\citeauthoryear{Popham, Woosley, \& Fryer}{1999}]{1999ApJ...518..356P} Popham R.,
    Woosley S.~E., Fryer C., 1999, ApJ, 518, 356


\bibitem[\protect\citeauthoryear{Ravi \& Lasky}{2014}]{2014MNRAS.441.2433R} Ravi V., Lasky P.~D., 2014,
    MNRAS, 441, 2433


\bibitem[\protect\citeauthoryear{Rosswog, Ramirez-Ruiz, \& Davies}{2003}]{2003MNRAS.345.1077R} Rosswog
    S., Ramirez-Ruiz E., Davies M.~B., 2003, MNRAS, 345, 1077


\bibitem[\protect\citeauthoryear{Rowlinson et al.}{2013}]{2013MNRAS.430.1061R} Rowlinson A., O'Brien
    P.~T., Metzger B.~D., Tanvir N.~R., Levan A.~J., 2013, MNRAS, 430, 1061


\bibitem[\protect\citeauthoryear{Rowlinson et al.}{2010}]{2010MNRAS.409..531R} Rowlinson A., et al.,
    2010, MNRAS, 409, 531


\bibitem[\protect\citeauthoryear{Sari, Piran, \& Narayan}{1998}]{1998ApJ...497L..17S} Sari R., Piran
    T., Narayan R., 1998, ApJ, 497, L17


\bibitem[\protect\citeauthoryear{Shapiro \& Teukolsky}{1983}]{1983bhwd.book.....S} Shapiro S.~L.,
    Teukolsky S.~A., 1983, bhwd.book,


\bibitem[\protect\citeauthoryear{Siegel \& Ciolfi}{2016}]{2016ApJ...819...15S} Siegel D.~M., Ciolfi R.,
    2016, ApJ, 819, 15


\bibitem[\protect\citeauthoryear{Thompson}{1994}]{1994MNRAS.270..480T} Thompson C., 1994, MNRAS, 270,
    480


\bibitem[\protect\citeauthoryear{Troja et al.}{2007}]{2007ApJ...665..599T} Troja E., et al., 2007, ApJ,
    665, 599


\bibitem[\protect\citeauthoryear{Troja et al.}{2018}]{2018MNRAS.478L..18T} Troja E., et al., 2018,
    MNRAS, 478, L18


\bibitem[\protect\citeauthoryear{Troja et al.}{2017}]{2017Natur.551...71T} Troja E., et al., 2017,
    Natur, 551, 71


\bibitem[\protect\citeauthoryear{Uhm \& Beloborodov}{2007}]{2007ApJ...665L..93U} Uhm Z.~L., Beloborodov
    A.~M., 2007, ApJ, 665, L93


\bibitem[\protect\citeauthoryear{Usov}{1992}]{1992Natur.357..472U} Usov V.~V., 1992, Natur, 357, 472


\bibitem[\protect\citeauthoryear{Wheeler et al.}{2000}]{2000ApJ...537..810W} Wheeler J.~C., Yi I.,
    H{\"o}flich P., Wang L., 2000, ApJ, 537, 810


\bibitem[\protect\citeauthoryear{Yu, Cheng, \& Cao}{2010}]{2010ApJ...715..477Y} Yu Y.-W., Cheng K.~S.,
    Cao X.-F., 2010, ApJ, 715, 477


\bibitem[\protect\citeauthoryear{Yu, Li, \& Dai}{2015}]{2015ApJ...806L...6Y} Yu Y.-W., Li S.-Z., Dai
    Z.-G., 2015, ApJ, 806, L6


\bibitem[\protect\citeauthoryear{Zhang}{2013}]{2013ApJ...763L..22Z} Zhang B., 2013, ApJ, 763, L22


\bibitem[\protect\citeauthoryear{Zhang}{2006}]{2006Natur.444.1010Z} Zhang B., 2006, Natur, 444, 1010


\bibitem[\protect\citeauthoryear{Zhang et al.}{2006}]{2006ApJ...642..354Z} Zhang B., Fan Y.~Z., Dyks
    J., Kobayashi S., M{\'e}sz{\'a}ros P., Burrows D.~N., Nousek J.~A., Gehrels N., 2006, ApJ, 642, 354


\bibitem[\protect\citeauthoryear{Zhang \& M{\'e}sz{\'a}ros}{2001}]{2001ApJ...552L..35Z} Zhang B.,
    M{\'e}sz{\'a}ros P., 2001, ApJ, 552, L35


\bibitem[\protect\citeauthoryear{Zhang et al.}{2007}]{2007ApJ...655L..25Z} Zhang B., Zhang B.-B., Liang
    E.-W., Gehrels N., Burrows D.~N., M{\'e}sz{\'a}ros P., 2007, ApJ, 655, L25


\bibitem[\protect\citeauthoryear{Zou et al.}{2018}]{2018 submitted} Zou L., Liang E.-W., L{\"u}
    H.-J., et al., 2018 submitted.

\end{thebibliography}
\end{document}